\documentclass[preprint,showpacs,preprintnumbers,amsmath,amssymb,
superscriptaddress]{revtex4}

\usepackage{graphicx}
\usepackage{dcolumn}
\usepackage{bm}

\begin{document}


\title{Investigation of double beta decay with the NEMO-3 detector}

\author{A.S.~Barabash} \email{barabash@itep.ru}
\affiliation{Institute of Theoretical and Experimental Physics, B.\
Cheremushkinskaya 25, 117218 Moscow, Russia}
\author{V.B.~Brudanin}
\affiliation{Joint Institute for Nuclear Research, 141980 Dubna, Russia}
\author{NEMO Collaboration}


\begin{abstract}
The double beta decay experiment NEMO~3 has been taking data since February 2003. The aim of this 
experiment is to search for neutrinoless ($0\nu\beta\beta$) decay and investigate two neutrino 
double beta decay in seven different isotopically 
enriched samples ($^{100}$Mo, $^{82}$Se, $^{48}$Ca, $^{96}$Zr, $^{116}$Cd, $^{130}$Te and $^{150}$Nd).
After analysis of the data corresponding to 3.75 y, no evidence for $0\nu\beta\beta$ decay 
in the $^{100}$Mo and $^{82}$Se samples was found. The half-life limits at the 90$\%$ C.L. 
are $1.1\cdot 10^{24}$ y and $3.6\cdot 10^{23}$ y, 
respectively. 
Additionally for $0\nu\beta\beta$ decay the following limits at the 90$\%$ C.L. were obtained, 
$> 1.3 \cdot 10^{22}$ y for $^{48}$Ca, $> 9.2 \cdot 10^{21}$ y for $^{96}$Zr and 
$> 1.8 \cdot 10^{22}$ y for $^{150}$Nd. The $2\nu\beta\beta$ decay half-life values were 
precisely measured for all investigated isotopes.   
\end{abstract}



\maketitle

\section{Introduction}

Interest in neutrinoless double-beta decay has seen a significant renewal in 
recent years after evidence for neutrino oscillations was obtained from the 
results of atmospheric, solar, reactor and accelerator  neutrino 
experiments (see, for example, the discussions in \cite{VAL06,BIL06,MOH06}). 
These results are impressive proof that neutrinos have a non-zero mass. However,
the experiments studying neutrino oscillations are not sensitive to the nature
of the neutrino mass (Dirac or Majorana) and provide no information on the 
absolute scale of the neutrino masses, since such experiments are sensitive 
only to the difference of the masses, $\Delta m^2$. The detection and study 
of $0\nu\beta\beta$ decay may clarify the following problems of neutrino 
physics (see discussions in \cite{PAS03,MOH05,PAS06}):
 (i) lepton number non-conservation, (ii) the nature of the 
neutrino (Dirac or Majorana particle), (iii) absolute neutrino
 mass scale (a measurement or a limit on $m_1$), (iv) the type of neutrino 
mass hierarchy (normal, inverted, or quasidegenerate), (v) CP violation in 
the lepton sector (measurement of the Majorana CP-violating phases).

The currently running NEMO~3 experiment is devoted to the search for $0\nu\beta\beta$ decay 
and to the accurate measurement of two neutrino double beta decay ($2\nu\beta\beta$ decay) by 
means of the direct detection of the two electrons. This tracking experiment, in contrast to experiments
with $^{76}$Ge, detects not only the total energy deposition,
but other parameters of the process. These include the energy of
the individual electrons, angle between them,
and the coordinates of the event in the source plane. Since June of 2002, the NEMO~3 detector
has operated in the Fr\'ejus Underground Laboratory (France)
located at a depth of 4800 m w.e. Then in February 2003, after the final tuning of the 
experimental set-up, NEMO~3 has been taking data devoted to double beta decay studies.
The first results with $^{100}$Mo, $^{82}$Se, $^{150}$Nd and $^{96}$Zr 
were published in \cite{ARN04,ARN05,ARN06,ARN07,ARG09,ARG09a}.

\section{NEMO-3 experiment}

The NEMO~3 detector  has three main components, a foil consisting
of different sources of double beta decay isotopes and copper, 
a tracker
made of Geiger wire cells and a calorimeter made of scintillator
blocks with PMT readout, surrounded by a solenoidal coil. The
detector has the ability to discriminate between events of different
types by positive identification of charged tracks and photons.  
A schematic view of the NEMO~3 detector is
shown in Fig. 1. 

\begin{figure*}
\begin{center}
\resizebox{0.5\textwidth}{!}{\includegraphics{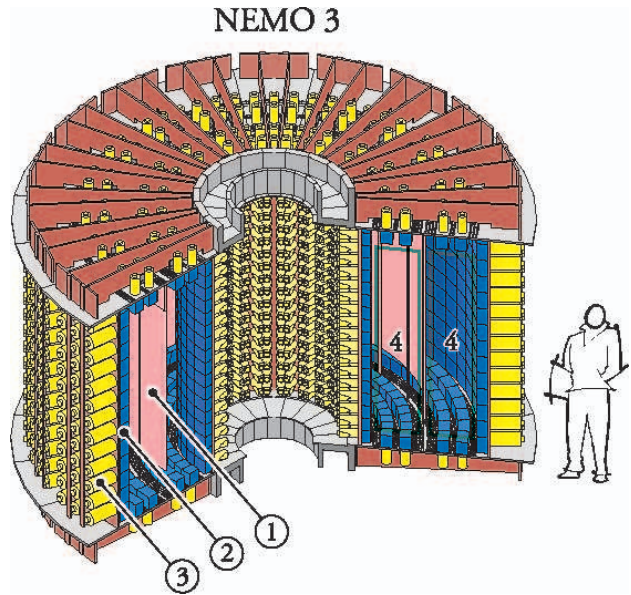}}
\caption{ The NEMO-3 detector without shielding. 1 -- source foil;
2-- plastic scintillator; 3 -- low radioactivity PMT; 4 --
tracking chamber.}
\label{fig1}
\end{center}
\end{figure*}

The  NEMO~3 detector is cylindrical in  design and is composed of
twenty equal sectors.  The external dimensions of the detector with
shields are 6 m in diameter and 4 m in height.  
NEMO~3 is based on the techniques tested on previous incarnations of
the experiment NEMO~1 \cite{DAS91}, and NEMO~2 \cite{ARN95}. 

The  wire chamber is  made of  6180 open  octagonal drift  cells which
operate in Geiger  mode (Geiger cells).  A gas  mixture of $\sim$ 95\%
helium, 4\% ethyl-alcohol, 1\% argon and 0.15\% water at 10 mbar above
atmospheric pressure is  used as the filling gas  of the wire chamber.
Each  drift  cell provides  a  three-dimensional  measurement of  the
charged particle tracks by recording the drift time and the two plasma
propagation  times.  The  transverse position  is determined  from the
drift  time,  while the  longitudinal  position  is  deduced from  the
difference between  the plasma propagation  times at both ends  of the
cathode wires.   The average vertex  position resolution for  the two-electron
events is $\sigma_t = 0.5$ cm  in the transverse plane of the detector
and $\sigma_l = 0.8$ cm in the longitudinal plane.
The Geiger cell information is treated by the track reconstruction
program   based  on  the   cellular  automaton   algorithm,  described
in \cite{KIS97}. 

The calorimeter, which surrounds the wire chamber, is composed of 1940
plastic   scintillator  blocks   coupled  by   light-guides   to  very
low-radioactivity   photomultiplier    tubes   (PMTs)   developed   by
Hammamatsu.   The energy resolution FWHM  of the calorimeter
ranges  from 14.1 to 17.6\% for  1 MeV  electrons, while  the  time
resolution is 250 ps at 1 MeV.

The apparatus accommodates almost 10 kg of different double beta decay
isotopes (see Table~1).
Most of  these  isotopes are highly enriched and all are
shaped in the  form  of thin  metallic  or composite  foils with  a
density of 30-60 mg/cm$^2$. 
Three sectors  are
used for external background  measurements and are equipped respectively
with pure Cu (one sector, 621 g) and natural Te (1.7 sectors, 614 g
of $^{nat}$TeO$_2$).  
Some of the sources, including $^{100}$Mo,
have been purified in
order  to  reduce their  content  of  $^{208}$Tl (from $^{232}$Th
and $^{228}$Th, and from $^{228}$Ra with a half-life of 5.75 y) and
$^{214}$Bi (from $^{226}$Ra with a half-life of 1600 y)
either  by  a  chemical  procedure \cite{ARN01},  or  by  a  physical
procedure \cite{ARN05a}.  The foils  were placed inside the wire chamber
in the  central vertical  plane of each  sector.  The majority  of the
detector, 12 sectors, notes a total of 6.9 kg of $^{100}$Mo.

\begin{table}
\caption{Investigated isotopes with NEMO~3 \cite{ARN05a}}
\vspace{0.5cm}
\begin{center}
\begin{tabular}{c|c|c|c|c|c|c|c}
\hline
Isotope & $^{100}$Mo & $^{82}$Se & $^{130}$Te & $^{116}$Cd &
$^{150}$Nd & $^{96}$Zr & $^{48}$Ca  \\
\hline
Enrichment, & 97 & 97 & 89 & 93 & 91 & 57 & 73 \\
\% & & & & & & & \\
Mass of & 6914 & 932 & 454 & 405 & 36.6 & 9.4 & 7.0 \\
isotope, g & & & & & & & \\
\hline
\end{tabular}
\end{center}
\end{table}

The detector is surrounded  by a solenoidal coil which generates
a vertical magnetic  field of 25 Gauss inside the  wire chamber.  This
magnetic field  allows electron-positron identification by measuring
the curvature of their tracks.  The ambiguity of the e$^+$/e$^-$ recognition
based on the curvature reconstruction is 3\% at 1 MeV.
 
The  whole detector  is  covered  by two  types  of shielding  against
external  $\gamma$-rays and  neutrons.  The  inner shield  is  made of
20 cm thick  low-radioactivity iron which stops $\gamma$-rays  and slow
neutrons.  The  outer shield is comprised  of tanks filled  with borated
water  on the  vertical walls  and  wood on  the top  and bottom 
designed to thermalize and capture neutrons.

At the beginning of the experiment, the radon inside the tracking chamber, 
and more precisely its decay product, $^{214}$Bi present in its radioactive 
chain, was found to be the predominant background. Radon is present 
in the air of the laboratory and originates from the rock surrounding. It 
can penetrate the detector through small leaks. A tent coupled 
to a radon-free air factory was installed around the detector in October 
2004 in order to decrease the presence of radon inside the tracker.

Since February 2003, after the final tuning of the
experimental set-up, NEMO~3 has  routinely been taking data devoted to
double beta  decay studies.  The calibration  with radioactive sources
is carried  out every  six weeks.  The  stability of the  calorimeter is
checked daily with a laser based calibration system \cite{ARN05a}.

The advantage of the NEMO~3 detector rests in its capability to
identify the two electrons from $\beta\beta$ decay and the
de-excitation photons from the  excited state of the daughter nucleus.
The  NEMO~3  calorimeter  also  measures  the detection  time  of  the
particles. The  use of appropriate time-of-flight (TOF) cuts,  in
addition to energy cuts, allows for an efficient reduction of all
backgrounds.

A full description of the detector and its characteristics can be found in \cite{ARN05a}.

\section{Experimental results}

The $\beta\beta$ events are selected by requiring two reconstructed
electron tracks with a curvature corresponding to a negative charge and
originating from a common vertex in the source foil.
The energy of each electron measured in the calorimeter should be greater than 200 keV.
Each track must hit a separate scintillator block.
No trackless PMT signals are allowed.
The event is recognized as internal using the measured time difference of the two PMT 
signals compared to the estimated time-of-flight difference of the electrons.

The background can be classified into three groups: external $\gamma$-rays,
radon inside the tracking volume and internal radioactive contamination of the source.
All three were estimated from NEMO~3 data with events of various topologies ~\cite{ARN09}.
In particular the radon and internal 
$^{214}$Bi is measured with e$\gamma$$\alpha$ events. 
The e$\gamma$, e$\gamma$$\gamma$ and e$\gamma$$\gamma$$\gamma$ events
are used to measure the $^{208}$Tl activity requiring
the detection of the 2.65 MeV $\gamma$-ray typical of the $^{208}$Tl $\beta$-decay.
Single electron events are used to measure the foil contamination by $\beta$-emitters.
The external background is measured with the events with the detected incoming $\gamma$-rays 
that produce 
an electron in the source foil. The control of the external background is done
with two-electron events
originating from the pure Copper and natural Tellurium foils.
Measurements performed with HPGe detectors and with radon detectors are used to verify the
results.
Different level of radon activity in the tracking volume were detected in the data taken before
October 2004 (Phase~1), and after the installation of the antiradon facility (Phase~2).

\subsection{Measurement of $2\nu\beta\beta$ half-lives}

Measurements of the $2\nu\beta\beta$ decay half-lives have been performed for seven isotopes
available in NEMO~3 (see Table 1). 
The NEMO~3  results of $2\nu\beta\beta$ half-life measurements are given 
in Table 2. For all the isotopes the energy sum spectrum, single-electron energy spectrum 
and angular distribution were measured.
The $^{100}$Mo double beta decay to the  $0^+_1$ excited state of $^{100}$Ru 
$T_{1/2}^{2\nu} = [5.7 ^{+1.3}_{-0.9}(stat) \pm 0.8 (syst)]\cdot10^{20}y$ 
has also been measured by NEMO 3 \cite{ARN07}. For $^{100}$Mo, $^{82}$Se, $^{96}$Zr and $^{150}$Nd 
these results are published.
For the other isotopes their status is preliminary.

\begin{table}
\caption{Two neutrino half-life values for different nuclei
obtained in
the NEMO~3 experiment (for $^{116}$Cd, $^{48}$Ca and $^{130}$Te 
the results are preliminary). 
The first error is statistical and the second is systematic while 
{\it S}/{\it B} is the signal-to-background ratio.}
\vspace{0.5cm}
\begin{center}
\begin{tabular*}{\textwidth}{l|@{\extracolsep{\fill}}c|c|c|c}
\hline
Isotope & Measurement & Number of & {\it S}/{\it B} & $T_{1/2}(2\nu)$, y \\
& time, days & $2\nu$ events & &  \\
\hline
$^{100}$Mo & 389 & 219000 & 40 & $(7.11 \pm 0.02 \pm
0.54 )\times 10^{18}$ \cite{ARN05}  \\
$^{100}$Mo-$^{100}$Ru($0^+_1$) & 334.3 & 37.5 & 4 & $(5.7^{+1.3}_{-0.9} \pm
0.8 )\times 10^{20}$ \cite{ARN07}  \\
$^{82}$Se & 389 & 2750 & 4 & $(9.6 \pm 0.3 \pm 1.0)
\times 10^{19}$ \cite{ARN05} \\
$^{116}$Cd & 168.4 & 1371 & 7.5 & $(2.8 \pm 0.1 \pm
0.3)\times 10^{19}$  \\
$^{96}$Zr & 1221 & 428 & 1 & $(2.35 \pm 0.14 \pm 0.19)
\times 10^{19}$ \cite{ARG09a} \\
$^{150}$Nd & 939 & 2018 & 2.8 & $(9.11^{+0.25}_{-0.22} \pm 0.63)
\times 10^{18}$ \cite{ARG09} \\
$^{48}$Ca & 943.16 & 116 & 6.8 & $(4.4^{+0.5}_{-0.4} \pm 0.4)
\times 10^{19}$  \\
$^{130}$Te & 1152 & 236 & 0.35 & $(6.9 \pm 0.9 \pm 1.0)
\times 10^{20}$  \\
\hline
\end{tabular*}
\end{center}
\end{table}

\subsection{Search for $0\nu\beta\beta$ decay}

No evidence for $0\nu\beta\beta$ decay was found for all seven isotopes. 
The associated limits are presented in Table 3.

\begin{table}[ht]
\label{Table6}
\caption{Limits at 90\% C.L. on $0\nu\beta\beta$ decay (neutrino mass mechanism) for different nuclei
obtained in
the NEMO~3 experiment.}
\vspace{0.5cm}
\begin{center}
\begin{tabular}{ccc}
\hline
Isotope & Measurement & $T_{1/2}(0\nu)$, y \\
& time, days &  \\
\hline
$^{100}$Mo & 1409 & $ > 1.1\cdot 10^{24}$ y  \\
$^{82}$Se & 1409 & $ > 3.6\cdot 10^{23}$ y  \\
$^{130}$Te & 1221 & $> 1 \cdot 10^{23}$ y   \\
$^{150}$Nd & 939 & $> 1.8\cdot 10^{22}$ y   \\
$^{116}$Cd & 77 & $> 1.6\cdot 10^{22}$ y   \\
$^{48}$Ca & 943.16 & $> 1.3\cdot 10^{22}$ y   \\
$^{96}$Zr & 1221 & $> 9.2\cdot 10^{21}$ y  \\

\hline
\end{tabular}
\end{center}
\end{table}

The $0\nu\beta\beta$-decay search in NEMO~3 is most promising with $^{100}$Mo and $^{82}$Se 
because of the larger available sample mass and high enough $Q_{\beta\beta} \sim$ 3 MeV.
The data taken from February 2003 till the end of 2008 has been used in the search for
neutrinoless double beta decay. The data corresponds to 1409 effective days of data collection.
For the case of the mass mechanism, the $0\nu\beta\beta$-decay signal is expected to be a peak 
in the energy sum distribution at the position of the transition energy $Q_{\beta\beta}$.
The two-electron energy sum spectra for $^{100}$Mo and $^{82}$Se are shown in Fig. 2 and Fig. 3
demonstrating a good agreement between observed and expected number of events.

\begin{figure}[htb]
\includegraphics[width=0.41\textwidth]{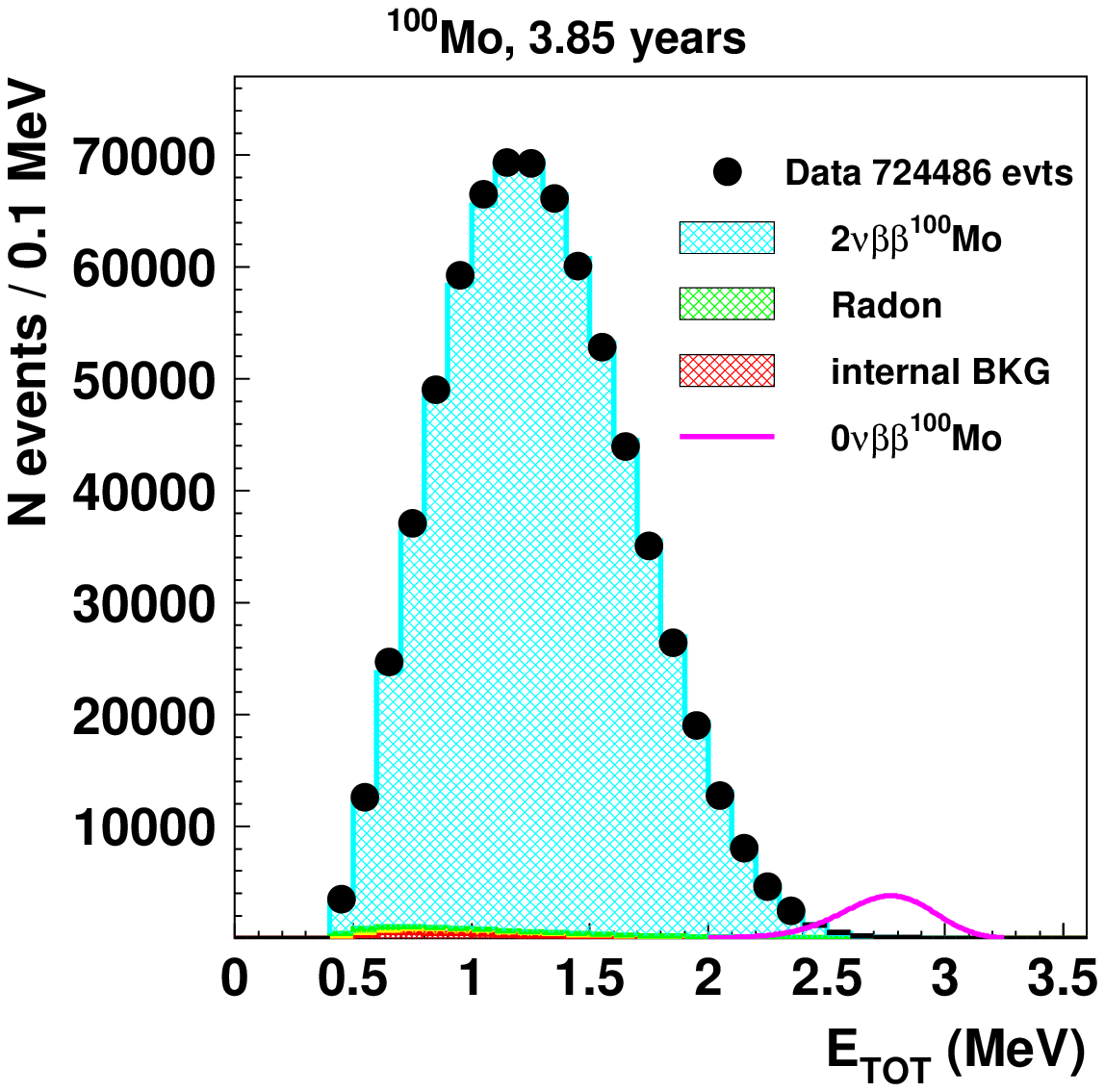}
\includegraphics[width=0.41\textwidth]{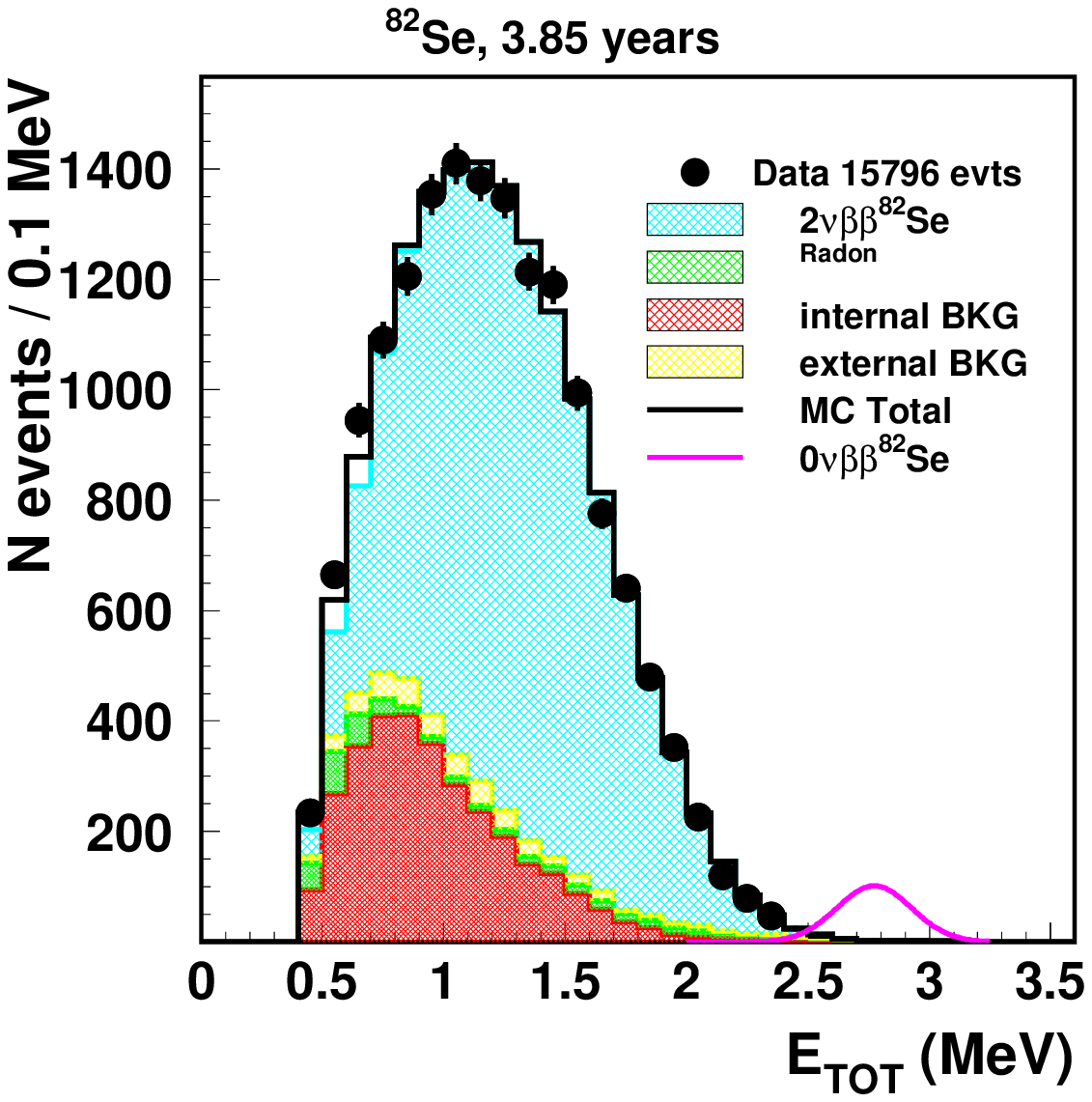}
\caption{Distribution of the energy sum of two electrons 
for $^{100}$Mo (left) and  $^{82}$Se (right), 1409 d data. The shape of a hypothetical 
$0\nu$ signal  is shown by the 
curve in arbitrary units.}
\end{figure}

\begin{figure}[htb]
\includegraphics[width=0.41\textwidth]{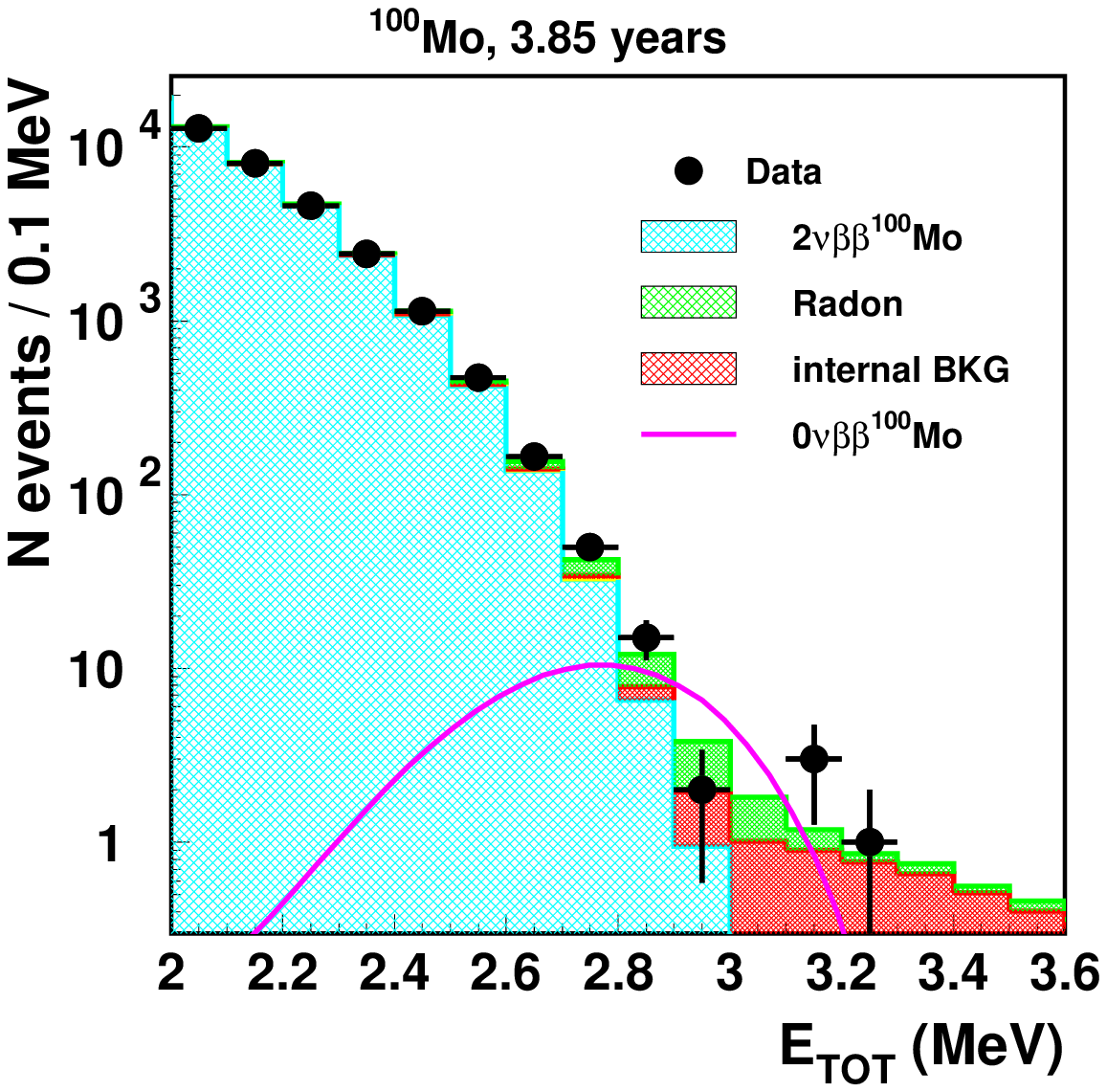}
\includegraphics[width=0.41\textwidth]{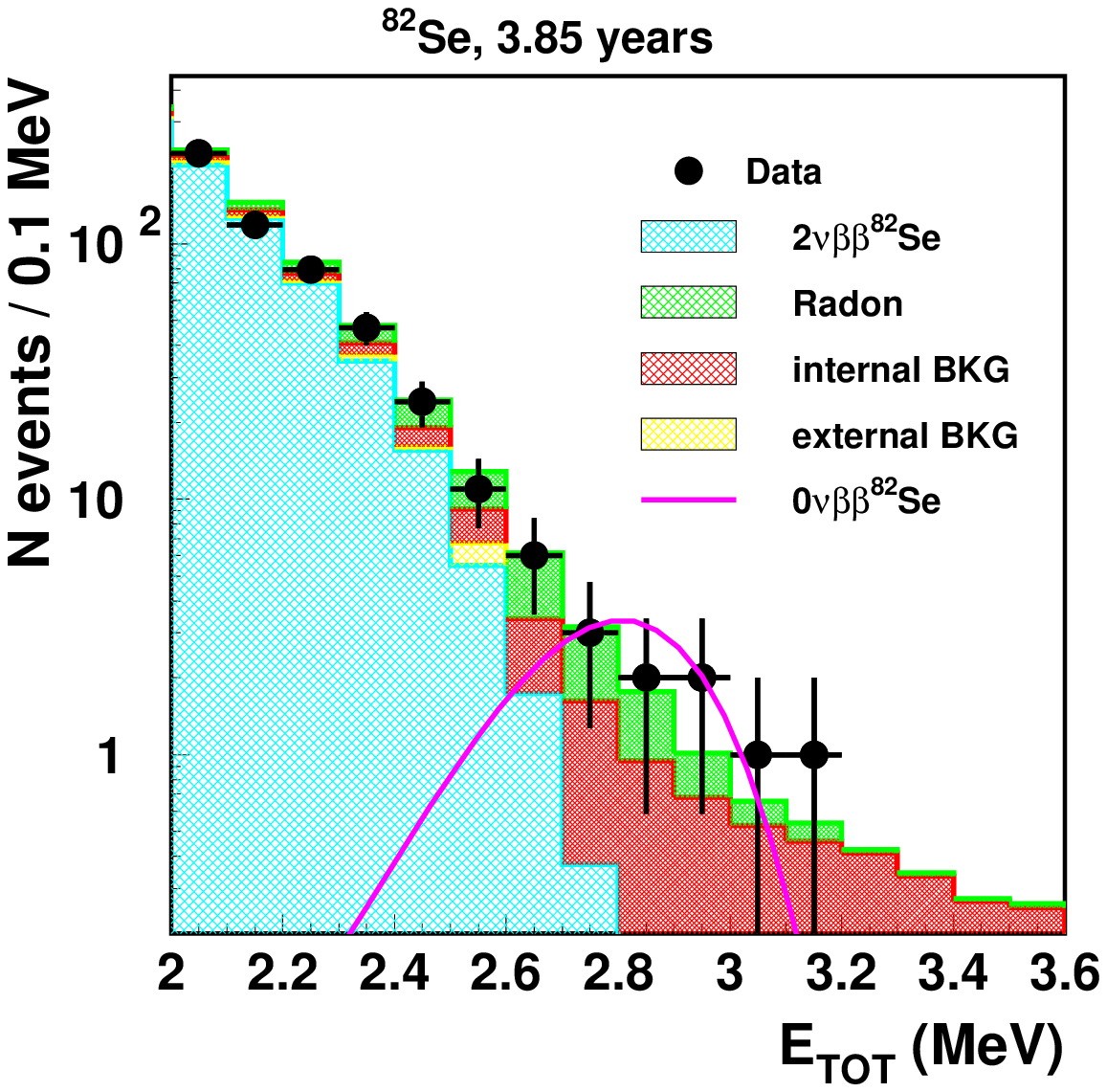}
\caption{Distribution of the energy sum of two electrons 
in the region around Q$_{\beta\beta}$ value 
for $^{100}$Mo (left) and  $^{82}$Se (right), 1409 d data.
High energy tail 
of the energy sum distribution for events in
molybdenum (left) and selenium (right) foils are shown with black
points. The background contributions are shown within the
histogram. The shape of a hypothetical $0\nu$ signal is shown by the 
curve in arbitrary units.}
\end{figure}

Since no excess is observed at the tail of the distributions,
limits are set on the neutrinoless double beta decay 
with the $CL_s$ method  based on the use of  a log-likelihood ratio (LLR) 
test statistics \cite{FIS07}.
This gives the lower limits on the half-life of 
T$_{1/2}^{0\nu}(^{100}Mo) > 1.1\cdot 10^{24} yr$ (90\% C.L.)
and 
T$_{1/2}^{0\nu}(^{82}Se) > 3.6\cdot 10^{23} yr$ (90\% C.L.)
A lower half-life limit translates into an 
upper limit on the effective Majorana neutrino mass $\langle m_{\nu}\rangle$.
According to the recent results of different theoretical NME calculations (see Table 4) the
half-life limit obtained for $^{100}$Mo corresponds to the neutrino mass interval
$\langle m_{\nu}\rangle <$  0.45 -- 0.93 eV. It is less restrictive for $^{82}$Se
$\langle m_{\nu}\rangle <$  0.89 -- 2.43 eV.
The reached NEMO 3 sensitivity on the neutrino mass is close to that of  IGEX~\cite{AAL02}, 
the Heidelberg-Moscow collaboration~\cite{KLA01} and CUORICINO~\cite{ARN08}.

\begin{table}[h]
\caption{Limits on the effective neutrino mass $\langle m_{\nu}\rangle$ (in eV) 
corresponding to different theoretical model calculations of nuclear matrix elements 
obtained for 
T$_{1/2}(0\nu\beta\beta) > 1.1\cdot 10^{24}$ y in case of $^{100}$Mo
and T$_{1/2}(0\nu\beta\beta) > 3.6\cdot 10^{23}$ y in case of $^{82}$Se.\label{tab:lim}}
\vspace{0.4cm}
\begin{tabular}{  l l c c }
\hline
Nuclear matrix elements & &$^{100}Mo$  & $^{82}Se$ \\
\hline
Shell model & \cite{CAU08}& - & $< 2.43$   \\
QRPA & \cite{KOR07},\cite{KOR07a}& $< 0.58 - 0.75$ & $< 1.12 - 1.38$  \\
QRPA & \cite{SIM08}& $< 0.45 - 0.93$ & $< 0.89 - 1.61$   \\
IBM-2 & \cite{IBM-2}& $< 0.49 - 0.55$ & $< 1.03 - 1.19$   \\
PHFB & \cite{PHFB}& $< 0.70$ &    \\
 \hline
\end{tabular}
\end{table}

\subsection{Search for double beta decay with Majoron emission}

The $0\nu\chi^{0}\beta\beta$ decay requires
the existence of a Majoron. It is a massless
Goldstone boson that arises due to a global breakdown of ($\it{B}$-$\it{L}$)
symmetry, where {\it B} and {\it L} are,
respectively, the baryon and the lepton number. The Majoron, if it
exists, could play a significant role
in the history of the early Universe and in the evolution of
stars. A
$2\beta$-decay model that involves the
emission of two Majorons was proposed within supersymmetric
theories and several other models of the
Majoron were proposed in the 1990s (see review \cite{BAR08} and references therein).
The possible two electrons energy spectra for different $0\nu\chi^{0}\beta\beta$
decay modes of $^{100}$Mo are shown
in Fig. 4. Here {\it n} is the spectral 
index, which defines
the shape of the spectrum. For example, for an ordinary Majoron {\it n} = 1, 
for 2$\nu$ decay {\it n} = 5, in the case of a bulk Majoron {\it n} = 2 
and for the process with two Majoron emission {\it n} = 3 or 7.

\begin{figure*}
\begin{center}
\resizebox{0.5\textwidth}{!}{\includegraphics{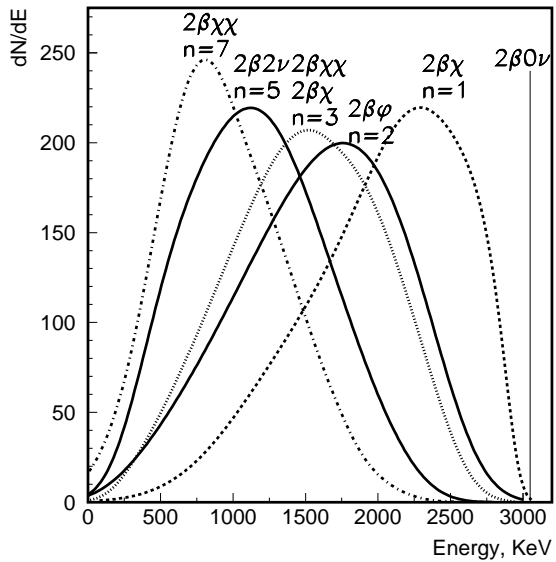}}
\caption{Energy spectra of different modes of $2\beta2\nu$ $(n=5)$,
$2\beta\chi^{0}$ $(n=1~$,$~2$ and$~3)$ and
$2\beta\chi^{0}\chi^{0} (n=3~$and$~7)$ decays of $~^{100}$Mo (see text).}
\label{fig_figure4}
\end{center}
\end{figure*}  

No evidence for $0\nu\chi^{0}\beta\beta$ decay was found for all seven isotopes. 
The limits for $^{100}$Mo, $^{82}$Se, $^{150}$Nd and $^{96}$Zr are presented in Table 5.
In particular, 
strong limits on "ordinary" Majoron 
(spectral index 1) decay of $^{100}\rm Mo$ ($T_{1/2} > 2.7\cdot10^{22}$ y) 
and $^{82}\rm Se$ ($T_{1/2} > 1.5\cdot10^{22}$ y)
have been obtained. Corresponding bounds on the Majoron-neutrino 
coupling constant are $<g_{ee}> < (0.35-0.85) \cdot 10^{-4}$ 
and  $< (0.6-1.9) \cdot 10^{-4}$, respectively (using nuclear matrix elements from 
\cite{CAU08,KOR07,KOR07a,SIM08,IBM-2,PHFB}).
\begin{table}
\caption{NEMO~3 limits on $T_{1/2}$ (y) for decay with one and
two Majorons at 90\% C.L. for modes
with spectral index {\it n} = 1, {\it n} = 2, {\it n} = 3 and {\it n} = 7.}
\vspace{0.5cm}
\begin{center}
\begin{tabular}{l|c|c|c|c}
\hline
Isotope & n =1 & n = 2 & n = 3 & n = 7 \\
\hline
$^{100}$Mo \cite{ARN05} & $>$2.7$\cdot$10$^{22}$ & $>$1.7$\cdot$10$^{22}$ & $>$1$\cdot$10$^{22}$ & 
$>$7$\cdot$10$^{19}$  \\
$^{82}$Se \cite{ARN05} & $>$1.5$\cdot$10$^{22}$ & $>$6$\cdot$10$^{21}$ & $>$3.1$\cdot$10$^{21}$ & 
$>$5$\cdot$10$^{20}$ \\
$^{96}$Zr \cite{ARG09a} & $>$1.9$\cdot$10$^{21}$ & $>$9.9$\cdot$10$^{20}$ & $>$5.8$\cdot$10$^{20}$ & 
$>$1.1$\cdot$10$^{20}$ \\
$^{150}$Nd \cite{ARG09} & $>$1.5$\cdot$10$^{21}$ & $>$5.4$\cdot$10$^{20}$ & $>$2.2$\cdot$10$^{20}$ & 
$>$4.7$\cdot$10$^{19}$ \\

\hline
\end{tabular}
\end{center}
\end{table}

\section{Conclusion}

The NEMO~3 detector has been operating within its target performance specifications since February 2003. 
The $2\nu\beta\beta$ decay  has been measured for seven isotopes with high statistics and greater 
precision than previously. The $^{100}$Mo $2\nu\beta\beta$ decay to the $0^+_1$ excited state of $^{100}$Ru 
has also been measured. 
No evidence for $0\nu\beta\beta$ decay was found for all seven isotopes. 
The best $0\nu\beta\beta$ limits have been obtained with $^{100}$Mo ($ > 1.1\cdot 10^{24}$ y at 90\% C.L.) 
and $^{82}$Se ($ > 3.6\cdot 10^{23}$ y at 90\% C.L.). 

NEMO~3 is current experiment with data collection continued into $\sim$ 2011.

\section*{Acknowledgements}

The authors thank the Modane Underground Laboratory staff for their 
technical 
assistance in running the experiment.  
This work was supported by grants from RFBR 
(no 06-02-72553 and no 09-02-92676)  and with support 
from the Russian Federal Agency for Atomic Energy.

\end{document}